\documentclass{article}

\usepackage{arxiv}

\usepackage[utf8]{inputenc} % allow utf-8 input
\usepackage[T1]{fontenc}    % use 8-bit T1 fonts
\usepackage{hyperref}       % hyperlinks
\usepackage{url}            % simple URL typesetting
\usepackage{booktabs}       % professional-quality tables
\usepackage{amsfonts}       % blackboard math symbols
\usepackage{nicefrac}       % compact symbols for 1/2, etc.
\usepackage{microtype}      % microtypography
\usepackage{lipsum}		% Can be removed after putting your text content
\usepackage{graphicx}
\usepackage{natbib}
\usepackage{doi}
\usepackage{caption}
\usepackage{subcaption}

\title{Self-Sensing Hysteresis-Type Bearingless Motor}

\author{ {\hspace{1mm}Laura Homiller}\\
	Walker Department of Mechanical Engineering\\
	University of Texas at Austin\\
	Austin, TX 78712 \\
	\texttt{laura.homiller@utexas.edu} \\
	\And
	{\hspace{1mm}Lei Zhou}\\
	Walker Department of Mechanical Engineering\\
	University of Texas at Austin\\
	Austin, TX 78712 \\
	\texttt{lzhou@utexas.edu} \\
}

\date{}

\renewcommand{\headeright}{}
\renewcommand{\undertitle}{}
\hypersetup{
pdftitle={Self-Sensing Hysteresis-Type Bearingless Motor},
pdfauthor={Laura Homiller, Lei Zhou},
}

\begin{document}
\maketitle
\begin{abstract}
	Bearingless motors use a single stator assembly to apply torque and magnetic suspension forces on the rotor, making these machines compact with frictionless operation and thus well suited to high-speed applications. One major challenge that prevents wide usage of bearingless motors is the need for air-gap position sensors, which are typically expensive. Here we present a method to estimate the radial position of a hysteresis-type bearingless motor using the inductance variation of the stator coils amplified by an injected high-frequency signal. We have carried out finite element (FE) simulations to demonstrate its feasibility, and have constructed a prototype self-sensing bearingless motor for experimental validations. 
\end{abstract}

\section{Introduction}
The development of high-speed machines is a growing trend in the electric machine industry. Increasing the speed decreases the volume and weight of the electric machine for the same power rating, which has advantages for electric transportation and portable power storage and generation applications.
\textit{Bearingless motors} are electric machines that apply torque and magnetic suspension forces on the rotor using a single stator assembly. These machines have demonstrated excellent performance for high-speed applications such as pumps  \citep{schoeb1997principle} and flywheels \citep{ooshima2010magnetic} due to their frictionless operation and imbalance elimination. Additionally, the combined functionality of motor and bearing in a single assembly makes these machines compact and attractive for use in small devices \citep{ooshima2010magnetic, puentener2018150}. One major challenge that prevents wide usage of bearingless motors is the need for air-gap position sensors. The magnetic suspension of the rotor in a bearingless motor requires closed-loop feedback control for stabilization \citep{chiba1994analysis}, but the required position sensors are typically expensive. If bearingless motors can operate without these sensors, system cost can be significantly reduced and these machines applied to a wider range of application areas. 

This extended abstract presents a method for hysteresis-type bearingless motors to estimate the rotor’s radial position without additional displacement sensors, aiming at enabling a new kind of self-sensing bearingless motor with lower cost and improved robustness. The key idea is to use the inductance variation of stator coils to estimate the rotor displacement, and use an injected high-frequency signal to amplify the sensitivity of winding impedance with respect to rotor displacement. We have carried out finite element (FE) simulations to demonstrate its feasibility, and have constructed a prototype self-sensing bearingless motor for experimental validations.

\begin{figure}[t!]
\centering
\begin{subfigure}[b]{0.4\textwidth}
\centering
\includegraphics[trim=160 50 170 40, clip, width =0.8\columnwidth, keepaspectratio=true]{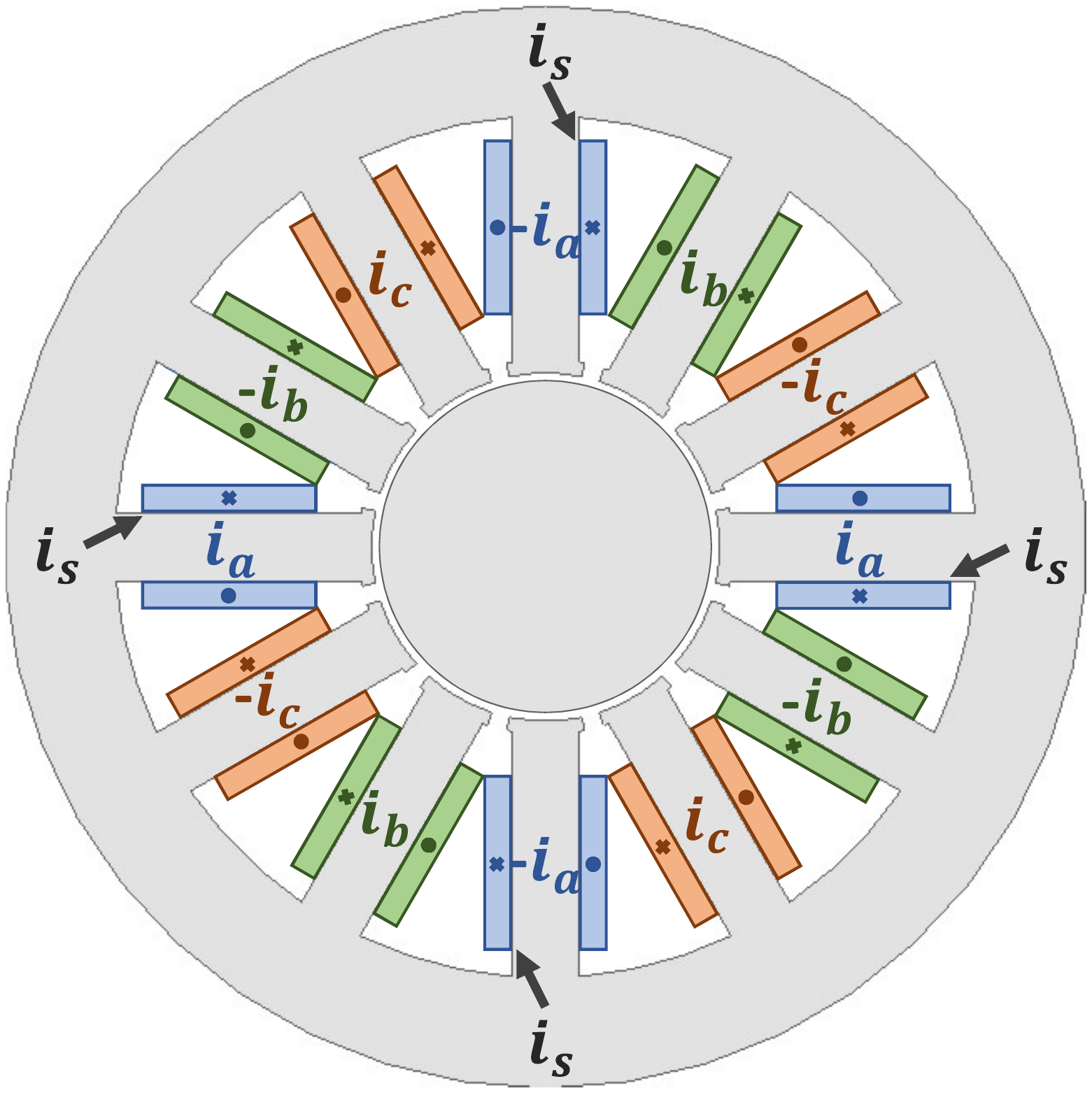}
\caption{}
\end{subfigure}
\begin{subfigure}[b]{0.4\textwidth}
\centering
\includegraphics[trim=160 50 170 40, clip,width =0.8\columnwidth, keepaspectratio=true]{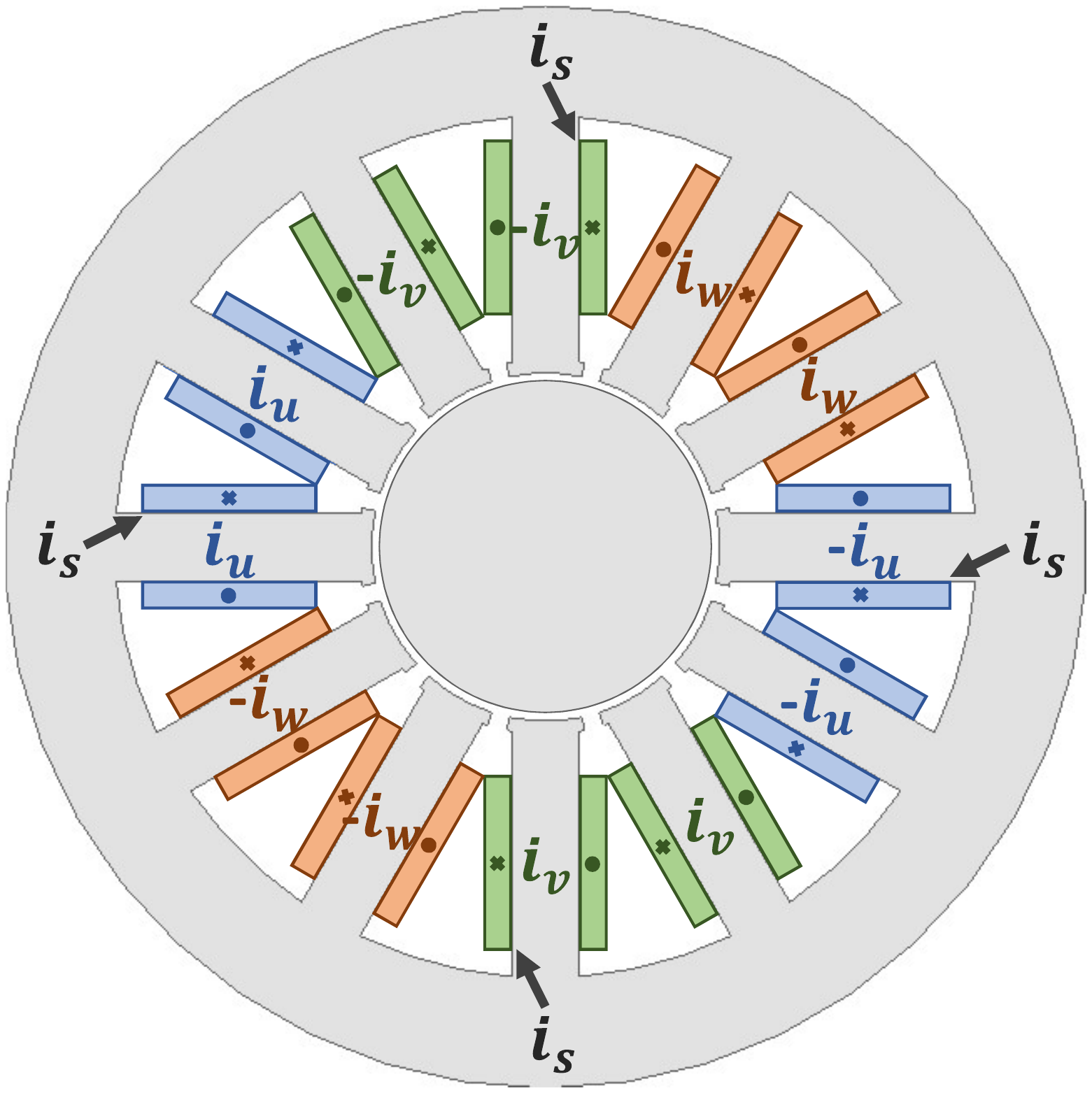}
\caption{}
\end{subfigure}
\caption{Winding pattern of prototype motor. $i_s$ is the high-frequency air gap displacement sensing current. (a) 4-pole motor winding. (b) 2-pole suspension control winding.  }
\label{fig:Winding_Pattern}
\end{figure}

\section{Operating Principle}

\underline{Bearingless Motor Operating Principle:}
The suspension control forces in a bearingless motor are generated by the superposition of two magnetic fields \cite{chiba1994analysis}. 
In our proposed motor design, a two-pole magnetic field is used for suspension, and a four-pole magnetic field is used for rotation.
Figure~\ref{fig:Winding_Pattern} shows the winding diagram of our proposed motor, where the stator has 12 teeth, and each of the stator coils wrap around one stator tooth and carry an independently-controlled current to generate both the two-pole and four-pole magnetic fields \citep{raggl2009comparison}. In this way, the stator generates both thrust torque and suspension controlling forces to the rotor simultaneously. 

In addition to the torque and force-generating currents, in order to enable the sensorless estimation for the air gap length, a high-frequency signal is injected to four stator coils's current as shown by $i_s$ in  Fig.~\ref{fig:Winding_Pattern}. For each lumped coil, the inductance can be calculated as
\begin{equation}
    L(g) = \frac{N^2\mu_0A}{2g},
\end{equation}
where $g$ is the air gap length at the coil's position, $N$ is the coil's number of turns, $\mu_0$ is the free-space magnetic permeability, and $A$ is the area of the stator tooth which the coil is wrapped around. 
The voltage across the coil is then related to the inductance of the coil according to: 
\begin{equation}
v = Ri + L\frac{di}{dt} + E_b,
\end{equation}
where $v$ and $i$ are the voltage and current in the coil, respectively, $R$ is the coil's resistance, and $E_b$ is the coil's back-electromotiveforce (back-EMF). In this work, since we are using a hysteresis motor, the back-EMF in the coils can be assumed to be small. Consequently, the voltage induced in each coil can be used to provide an estimate for the rotor's radial displacement.

\begin{figure}[t]
\centering
\includegraphics[trim=90 60 70 60, clip, width =0.45\columnwidth, keepaspectratio=true]{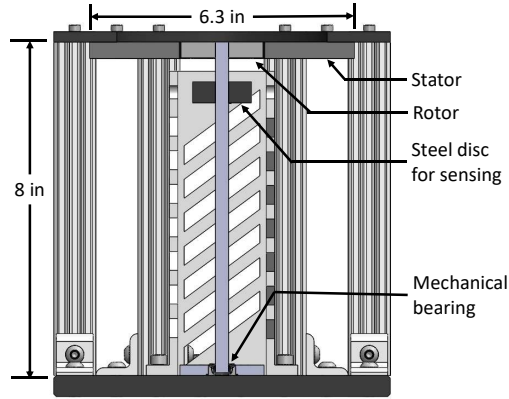}
\caption{Cross-section view for the CAD design of self-sensing bearingless hysteresis motor prototype. }
\label{fig:CAD_Model}
\end{figure}

\begin{figure}[t!]
\begin{centering}
\begin{subfigure}[b]{0.45\textwidth}
\centering
\includegraphics[trim=150 90 150 100, clip, width =0.9\columnwidth, keepaspectratio=true]{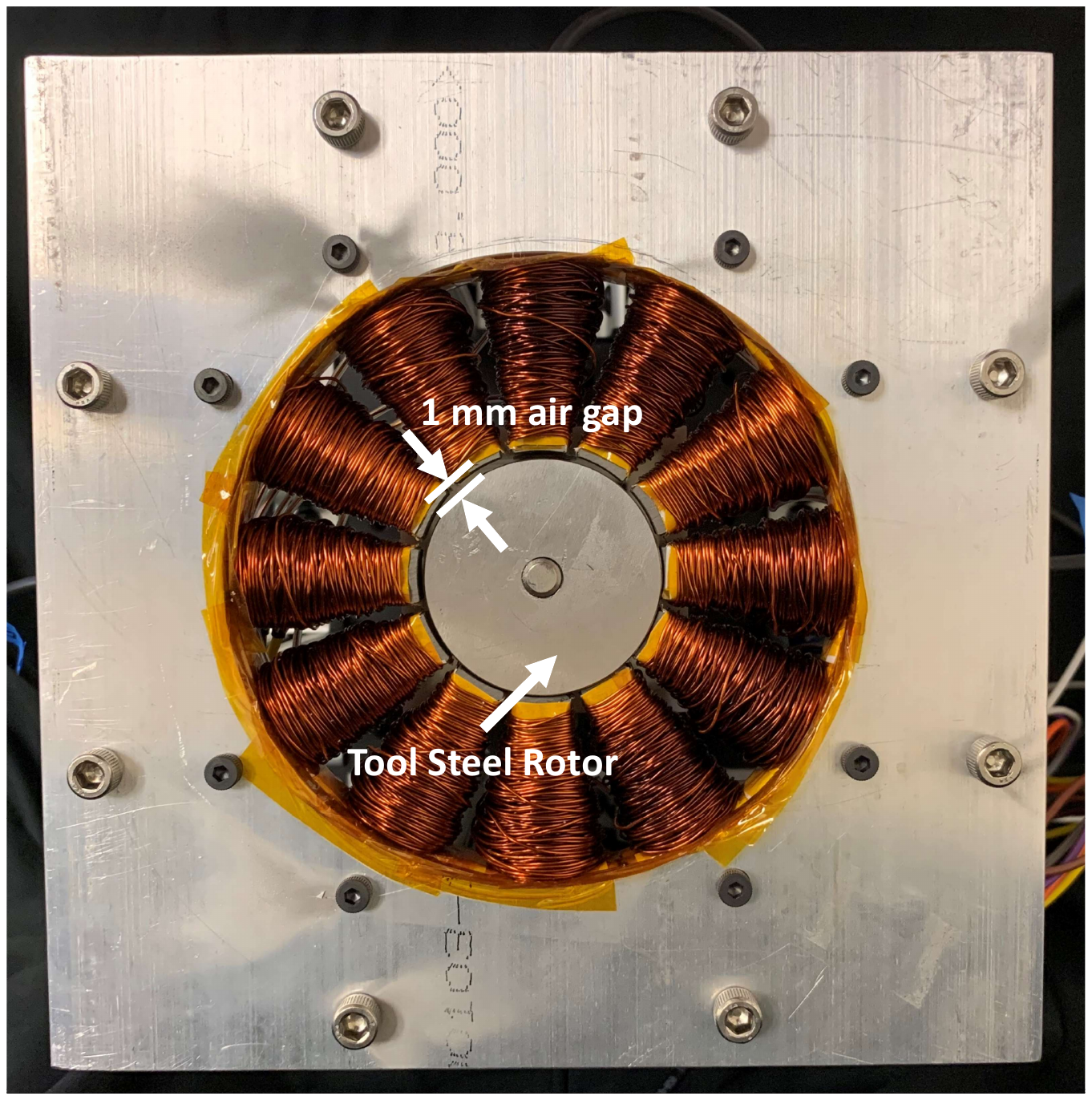}
\caption{}
\end{subfigure}
\begin{subfigure}[b]{0.45\textwidth}
\centering
\includegraphics[trim=170 100 170 100, clip,width =0.9\columnwidth, keepaspectratio=true]{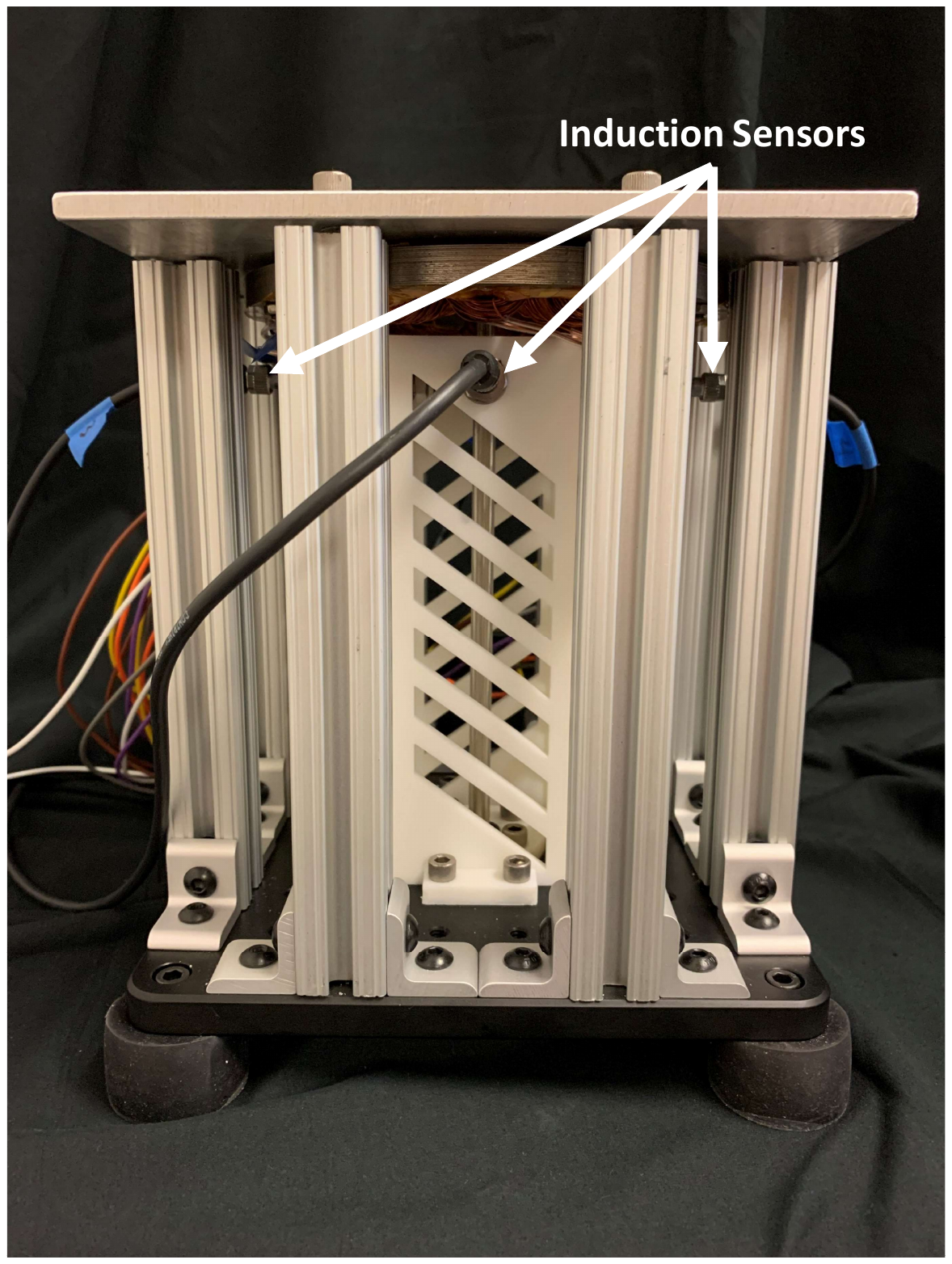}
\caption{}
\end{subfigure}
\caption{Photographs of motor prototype. (a) Top view showing the stator and rotor. (b) Side view of the structure showing sensors. }
\label{fig:motor_photo}
\end{centering}
\end{figure}

\begin{figure}[t]
\centering
\includegraphics[trim=0 0 40 30, clip, width =0.35\columnwidth, keepaspectratio=true]{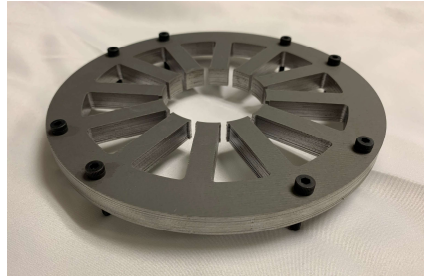}
\caption{Photograph of stator laminations. }
\label{fig:Stator_Picture}
\end{figure}

\begin{figure}[t]
\centering
\includegraphics[trim=0 150 0 150, clip, width =1\textwidth, keepaspectratio=true]{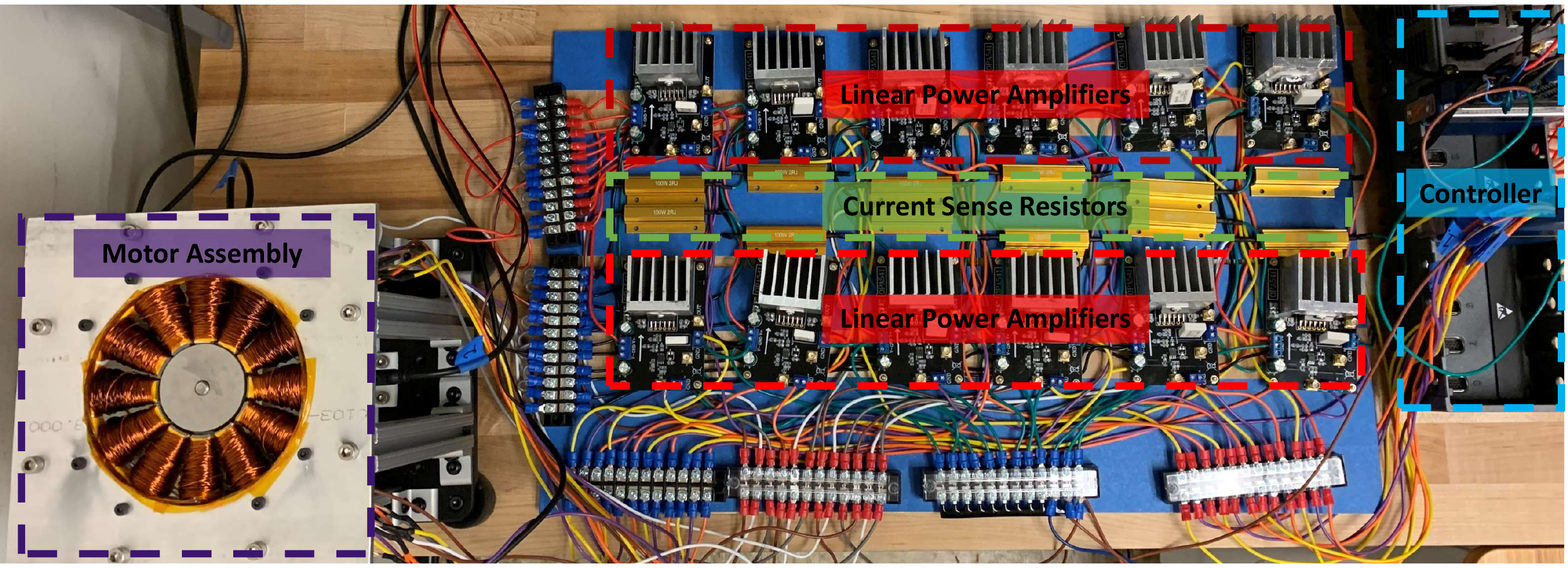}\\
\caption{Photograph of motor setup. Each coil is individually controlled using NI CompactRIO, and 12 linear power amplifiers amplify the signal sent to the coils.}
\label{fig:Full_Setup}
\end{figure}

\section{Hardware Overview}
We have designed a hysteresis-type bearingless motor testbed, and Fig.~\ref{fig:CAD_Model} shows a cross-section view of the CAD design. In this system,  the rotor is on top of a vertical shaft, where the suspension forces of the motor are used to constrain the radial position of the shaft on the top end. In addition, a self-aligning mechanical bearing is used to constrain the radial position of the shaft at the bottom end. The self-aligning bearing is selected to allow angular movement of the shaft so that the position of the rotor is not mechanically constrained. Four inductive sensors (DW-AS-509, Contrinex) measure the rotor’s radial position through a steel sensing disc mounted on the shaft, providing ``ground truth'' signals for the air gap length estimation problem. We will first test the magnetic suspension of the bearingless motor using the measured rotor's radial position, and the final system will use the estimated rotor position for the feedback control of the rotor’s magnetic suspension. 

We have constructed a prototype for the self-sensing bearingless motor testbed as shown in 
 Fig.~\ref{fig:motor_photo} and \ref{fig:Full_Setup}. In this setup, the rotor has a diameter of 49.5~mm and an axial length of 10~mm, and is made of D2 tool steel, which has large magnetic hysteresis and has been used in hysteresis motors as rotor materials \cite{zhou2018position}. Although hysteresis materials with better torque generation capability exist, in this work, we selected D2 tool steel as the rotor material due to its accessibility. The stator consists of lamination, insulation (Nomex 410 and KPT-1/2, DuPont), and winding coils. Figure~\ref{fig:Stator_Picture} shows a photograph of the stator lamination, which is made of 20 layers of M19 steel of 0.5~mm thickness. The stator also uses a deep-slot configuration to allow larger winding volume. 
 The stator coils are fabricated by magnetic wires of AWG22 (Epoxy Bondable MW, MWS). Each of the 12 stator coils
  is independently controlled and is driven by a linear power amplifier (OPA541, Texas Instruments). The coils are voltage controlled for initial testing, but analog current control will be implemented for future tests. The data acquisition of the sensor signals and realtime control uses NI compactRIO 9048. 

\begin{figure*}[t!]
\centering
\fbox{\includegraphics[trim=0 0 0 0, clip, width =0.95\textwidth, keepaspectratio=true]{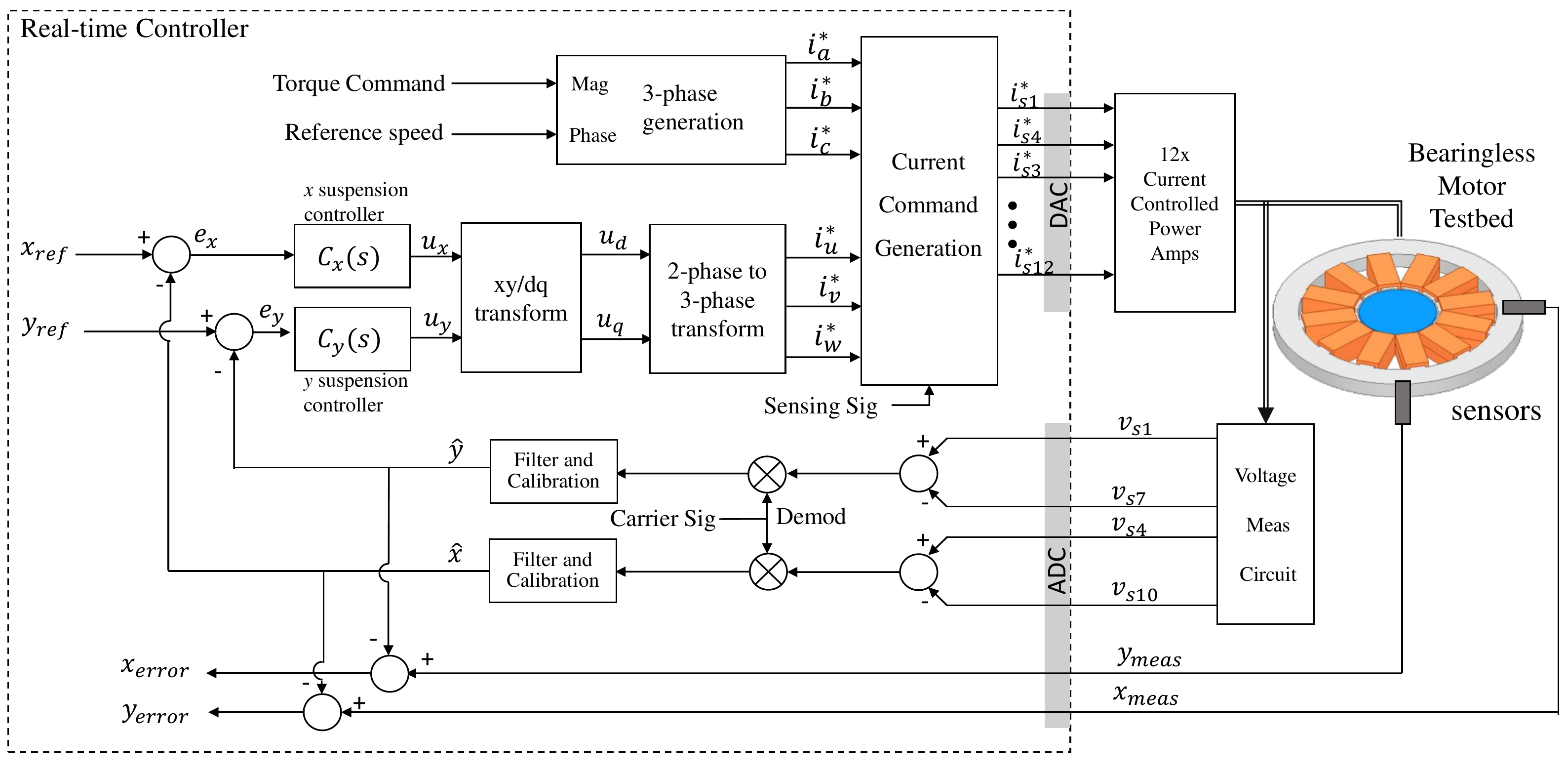}}\\
\caption{Control block diagram.}
\label{fig:Block_Diagram}
\end{figure*}

Figure~\ref{fig:Block_Diagram} shows a block diagram for the control system for the self-sensing bearingless motor. The difference between the voltages across coils 1 and 7, which are aligned with the positive and negative x-axis respectively, is multiplied by the high frequency carrier signal and then averaged using a moving average filter, givings an estimate for the x-directional displacement of the rotor $\hat{x}$. The estimate for the rotor's y-directional displacement $\hat{y}$ is calculated using the same process for the voltages across coils 4 and 10, which are located on the positive and negative y-axis. The estimate for rotor displacement is both compared with the displacements measured by the induction sensors $x_{meas}$ and $y_{meas}$ to give the error of the estimator and fed back for suspension control. The suspension controllers $C_x(s)$ and $C_y(s)$ take the error signals $e_x$ and $e_y$ and output the control efforts $u_x$ and $u_y$. These control efforts are transformed into $d$- and $q$-axis control efforts $u_d$ and $u_q$, and then transformed to 3-phase suspension current commands $i^*_u$, $i^*_v$, and $i^*_w$. The suspension current commands are superimposed with the torque current commands and high frequency sensing signal according to the winding diagram in Fig.~\ref{fig:Winding_Pattern}. These currents are passed to the power amplifiers, which drive the stator coils.

\section{Simulation Validation}
\begin{figure}[t!]
\centering
\includegraphics[width =0.6\columnwidth, keepaspectratio=true]{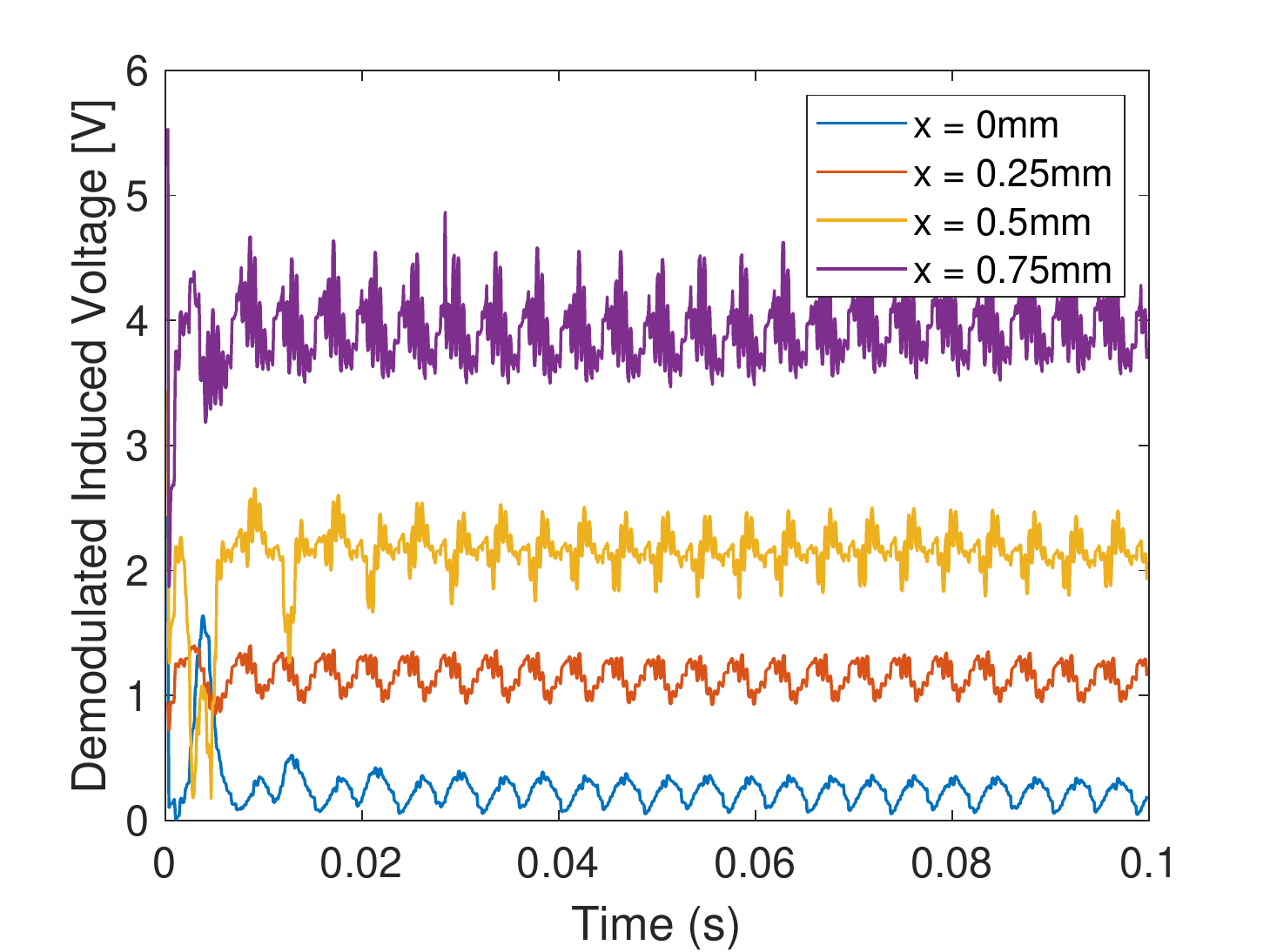}\\
\caption{Simulated induced voltage results for varied rotor position.  }
\label{fig:Ind_Voltages}
\end{figure}

We have conducted transient-time magnetic finite element simulations for the proposed air gap displacement sensing approach described in Figure~\ref{fig:Block_Diagram} using Ansys Maxwell. Fig.~\ref{fig:Ind_Voltages} shows simulated demodulated voltages of the proposed motor with injecting a sensing current of $0.1~\rm{A}$ at 2~kHz. The clear variation of signal with rotor radial displacement demonstrates the feasibility of the proposed approach. 

\section{Conclusion and Future Work}
The frictionless operation of bearingless motors makes the use of both the rotary and linear varieties of such machines attractive for a variety of precision applications. Eliminating the need for position sensors in such systems improves robustness while reducing complexity and cost. Finite element simulation has confirmed the feasibility of estimating the position of the rotor using the voltage induced in the stator coils, amplified by the injection of a high frequency signal. A prototype has been constructed to test this approach.

After implementing and tuning the position estimation for the current setup, there are several opportunities for improving the performance of this motor design. We plan to test and experimentally evaluate different winding configurations and solutions for the systems' power electronics.

\bibliographystyle{unsrtnat}
\bibliography{references}  

\end{document}